\documentclass[12pt]{iopart}

\usepackage[dvips]{graphicx}
\usepackage{array}
\usepackage{amssymb}
\usepackage{hhline}
\usepackage{longtable}
\usepackage{dcolumn}% Align table columns on decimal point
\usepackage{bm}% bold math
\usepackage{subfigure}
\usepackage[hyperindex,CJKbookmarks,dvipdfm,colorlinks,citecolor=blue,linkcolor=blue]{hyperref}
\begin{document}

\title{An alternative approach to determining average distance in a class of scale-free modular networks}

\author{Zhongzhi Zhang$^1$${}^,$$^2$, Yuan Lin$^1$${}^,$$^2$,  Shuigeng
Zhou$^1$${}^,$$^2$, Zhigang Wang$^3$, and Jihong Guan$^4$}

\address{$^1$ School of Computer Science, Fudan University,
Shanghai 200433, China}
\address{$^2$ Shanghai Key Lab of Intelligent Information
Processing, Fudan University, Shanghai 200433, China}
\address{$^3$ Mahle Technologies Holding (China) Co., Ltd., Shanghai 201400, China}
\address{$^4$ Department of Computer Science and Technology,
Tongji University, 4800 Cao'an Road, Shanghai 201804, China}

\eads{zhangzz@fudan.edu.cn, sgzhou@fudan.edu.cn,
jhguan@tongji.edu.cn}

\begin{abstract}
Various real-life networks of current interest are simultaneously
scale-free and modular. Here we study analytically the average
distance in a class of deterministically growing scale-free modular
networks. By virtue of the recursive relations derived from the
self-similar structure of the networks, we compute rigorously this
important quantity, obtaining an explicit closed-form solution,
which recovers the previous result and is corroborated by extensive
numerical calculations. The obtained exact expression shows that the
average distance scales logarithmically with the number of nodes in
the networks, indicating an existence of small-world behavior. We
present that this small-world phenomenon comes from the peculiar
architecture of the network family.
\end{abstract}

\pacs{89.75.Hc, 89.75.Da, 02.10.Ox, 05.10.-a}
%02.10.Ox Combinatorics; graph theory
%89.75.Da Systems obeying scaling laws
%89.75.Fb Structures and organization in complex systems
%89.75.Hc Networks and genealogical trees
%89.75.-k Complex systems
%05.10.-a Computational methods in statistical physics and nonlinear
%                dynamics
%99.10.Jk Corrected article

\maketitle

%\tableofcontents

\section{Introduction}

Average distance is one of the most important measurements
characterizing complex networks, which is a subject attracting a lot
of interest in the recent physics
literature~\cite{AlBa02,DoMe02,Ne03,BoLaMoChHw06}. Extensive
empirical studies showed that many, perhaps most, real networks
exhibit remarkable small-world phenomenon~\cite{WaSt98}, with their
average distance grows as a function of network order (i.e., number
of nodes in a network), or slowly~\cite{AlBa02,DoMe02}. As a
fundamental topological property, average distance is closely
related to other structural characteristics, such as degree
distribution~\cite{ChLu02,CoHa03}, centrality~\cite{DoMeOl06},
fractality~\cite{SoHaMa06, ZhZhZo07,ZhZhChGu08},
symmetry~\cite{XiMaiWaXiWa08}, and so forth. All these features
together play significant roles in characterizing and understanding
the complexity of networks. Moreover, average distance is relevant
to various dynamical processes occurring on complex networks,
including epidemic spreading~\cite{WaSt98}, target
search~\cite{AdLuPuHu01}, synchronization~\cite{NiMoLaHo03}, random
walks~\cite{CoBeTeVoKl07,ZhLiZhWuGu09, ZhQiZhGaGu09}, and many more.

In addition to the small-world behavior, other two prominent
properties that seem to be common to real networks, especially
biological and social networks, are scale-free feature~\cite{BaAl99}
and modular structure~\cite{GiNe02,RaSoMoOlBa02,RaBa03}. The former
implies that the networks obey a power-law degree distribution as
$P(k) \sim k^{-\gamma}$ with $2< \gamma \leq 3$, while the latter
means that the networks can be divided into groups (modules), within
which nodes are more tightly connected with each other than with
nodes outside. In order to describe simultaneously the two striking
properties, Ravasz and Barab\'asi (RB) presented a famous
model~\cite{RaBa03}, mimicking scale-free modular networks. Many
topological properties of and dynamical processes on the RB model
have been investigated in much detail, including degree
distribution~\cite{RaBa03}, clustering coefficient~\cite{RaBa03,
No03},  betweenness centrality distribution~\cite{No03}, community
structure~\cite{Zh03}, random walks~\cite{NoRi04a,ZhLiGaZhGuLi09},
among others. Particularly, by mapping the networks onto a Potts
model in one-dimensional lattices, Noh proved that the RB model is
small-world~\cite{No03}.

In this paper, we study the average distance in the RB model by
using an alternative approach very different from the previous
one~\cite{No03}. Our computation method is based on the particular
deterministic construction of the RB model. Concretely,  making use
of the self-similar structure of the scale-free modular networks, we
establish some recursion relations, from which we further derive the
exactly analytical solution to the average distance. Our obtained
rigorous expression is compatible with the previous formula. We show
that the RB model is small-world. We also show that the small-world
behavior is a natural result of the scale-free and modular
architecture of the networks under consideration.

\section{The modular scale-free networks}

\begin{figure}
\begin{center}
\includegraphics[width=0.50\linewidth,trim=80 220 80 160]{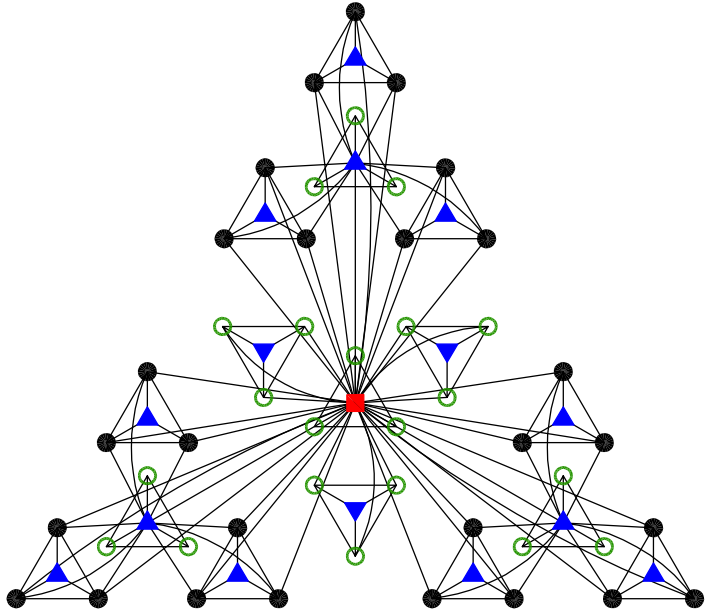}
\end{center}
\caption[kurzform]{(Color online) Sketch of a network $H_3$ for the
case of $m=4$. The filled squares and circles represent the hub node
and peripheral nodes, respectively.} \label{network}
\end{figure}

We first introduce the RB model for the scale-free modular networks,
which are built in an iterative way~\cite{RaSoMoOlBa02,RaBa03}. Let
$H_{g}$ stand for the network model after $g$ ($g\geq 1$) iterations
(i.e., number of generations). Initially ($g=1$), the model is
composed $m$ ($m \geq 3$) nodes linked by $m(m-1)/2$ edges forming a
complete graph, among which a node (e.g., the central node in
figure~\ref{network}) is called hub (or root) node, and the other
$m-1$ nodes are named peripheral nodes. At the second generation
($g=2$), $m-1$ replicas of $H_{1}$ are created with the $m-1$
peripheral nodes of each copy being connected to the root of the
original $H_{1}$. In this way, we obtained $H_{2}$, the hub and
peripheral nodes of which are the hub of the original $H_{1}$ and
the $(m-1)^2$ peripheral nodes in the $m-1$ duplicates of $H_{1}$,
respectively. Suppose one has $H_{g-1}$, the next generation network
$H_{g}$ can be obtained by adding $M-1$ copies of $H_{g-1}$ to the
primal $H_{g-1}$, with all peripheral nodes of the replicas being
linked to the hub of the original $H_{g-1}$ unit. The hub of the
original $H_{g-1}$ and the peripheral nodes of the $m-1$ copies of
$H_{g-1}$ form the hub node and peripheral nodes of $H_{g}$,
respectively. Repeating indefinitely the two steps of replication
and connection, one obtains the scale-free modular networks.
Figure~\ref{network} illustrates a network $H_3$ for the particular
case of $m=4$.

%%%%%%%%%%%%%%%%%%%%%%%%%%%%%%%%%%%%%%%%%%%%%%%%%%%%%%%%%
% Figure  1
%%%%%%%%%%%%%%%%%%%%%%%%%%%%%%%%%%%%%%%%%%%%%%%%%%%%%%%%%%
%%%%%%%%%%%%%%%%%%%%%%%%%%%%%%%%%%%%%%%%%%%%%%%%%%%%%%%%%%

Many interesting quantities of the model can be determined
explicitly~\cite{RaBa03,No03}. In $H_{g}$, the network order,
denoted by $N_g$ is $N_g=m^{g}$; the degree $K_h(g)=
\frac{m-1}{m-2}[(m-1)^g-1]$ of the hub node is the largest among all
nodes; the number of peripheral nodes, forming a set $\mathbb{P}_g$,
is $P_g=(m-1)^g $; and the average degree is approximately equal to
a constant $2(m-1)(3m-2)/m$ in the limit of infinite $g$, showing
that the networks are sparse.

The model under consideration is in fact an extension of the one
proposed in~\cite{BaRaVi01} and studied in much detail
in~\cite{IgYa05,AgBu09,ZhLiGaZhGu09}. It presents some typical
features observed in a variety of real-world
systems~\cite{RaBa03,No03}. Its degree distribution follows a
power-law scaling $P(k) \sim k^{-\gamma}$ with a general exponent
$\gamma=1+\ln m/\ln (m-1)$ belonging to the interval $(2,2.585)$.
Its average clustering coefficient tends to a large constant
dependent on $m$; and its average distance grows logarithmically
with the network order, both of which show that the model is
small-world. In addition, the betweenness distribution $P_B$ of
nodes also obeys the power-law behavior $P_B \sim B^{-2}$ with the
exponent regardless of the parameter $m$. Particularly, the whole
class of the networks shows a remarkable modular structure. These
peculiar structural properties make the networks unique within the
category of complex networks. %Many dynamical processes on these networks have been investigated~\cite{NoRi04a,ZhLiGaZhGuLi09}.

\section{Explicit formula for average distance}

As shown in the introduction section, average distance is closely
related to many topological properties of and various dynamical
processes on complex networks. In what follows, we will derive
analytically the average distance of the scale-free modular networks
by applying an alternative method completely different from that
in~\cite{No03}. We represent all the shortest path lengths of
network $H_{g}$ as a matrix in which the entry $d_{ij}(g)$ is the
distance between nodes $i$ and $j$ that is the length of a shortest
path joining $i$ and $j$. A measure of the typical separation
between two nodes in $H_{g}$ is given by the average distance
$d_{g}$ defined as the mean of distances over all pairs of nodes:
\begin{equation}\label{apl01}
d_{g}  = \frac{D_g}{N_g(N_g-1)/2}\,,
\end{equation}
where
\begin{equation}\label{total01}
D_g = \sum_{i \in H_{g},\, j \in H_{g},\, i \neq j} d_{ij}(g)
\end{equation}
denotes the sum of the distances between two nodes over all couples.
Notice that in Eq.~(\ref{total01}), for a pair of nodes $i$ and $j$
($i \neq j$), we only count $d_{ij}(g)$ or $d_{ji}(g)$, not both.

%%%%%%%%%%%%%%%%%%%%%%%%%%%%%%%%%%%%%%%%%%%%%%%%%%%%%%%%%
% Figure  2
%%%%%%%%%%%%%%%%%%%%%%%%%%%%%%%%%%%%%%%%%%%%%%%%%%%%%%%%%%
\begin{figure}
\begin{center}
\includegraphics[width=0.45\linewidth,trim=10 10 10 0]{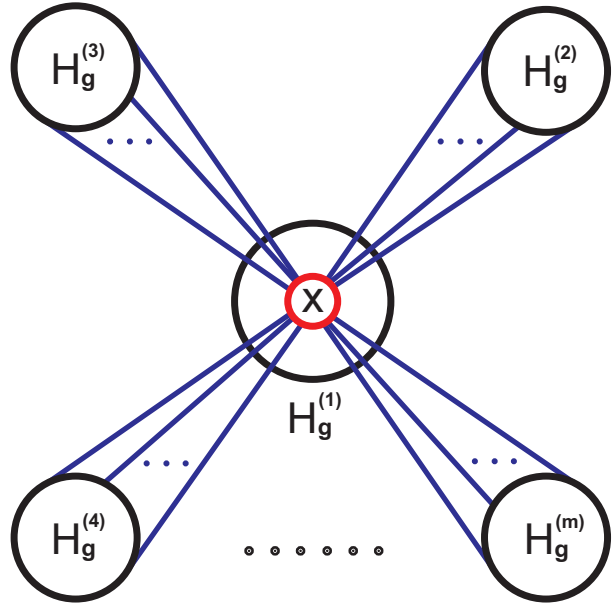}
\end{center}
\caption[kurzform]{\label{labeling} (Color online) Schematic
illustration of the means of construction of the scale-free modular
networks. $H_{g+1}$ is obtained by joining $m$ replicas of $H_{g}$
denoted as
    $H_{g}^{(\varphi)}$ $(\varphi=1,2, \cdots, m)$, which are
    connected to one another by linking all the peripheral nodes of $H_{g}^{(\chi)}$ ($2 \leq  \chi \leq m$) to the hub node (denoted by $X$) of $H_{g}^{(1)}$.}
\end{figure}
%%%%%%%%%%%%%%%%%%%%%%%%%%%%%%%%%%%%%%%%%%%%%%%%%%%%%%%%%%

We continue by exhibiting the procedure of determining the total
distance and present the recurrence formula, which allows us to
obtain $D_{g+1}$ of the $g+1$ generation from $D_{g}$ of the $g$
generation. The studied network $H_{g}$ has a self-similar structure
that allows one to calculate $D_{g}$ analytically. By construction
(see figure~\ref{labeling}), network $H_{g+1}$ is obtained by
joining $m$ copies of $H_{g}$ that are labeled as $H_{g}^{(1)}$,
$H_{g}^{(2)}$, $\cdots$, $H_{g}^{(m)}$. Using this self-similar
property, the total distance $D_{g+1}$ satisfies the recursion
relation
\begin{equation}\label{total02}
  D_{g+1} = m\, D_g + \Delta_g,
\end{equation}
where $\Delta_g$ is the sum over all shortest path length whose
endpoints are not in the same $H_{g}^{(\varphi)}$ branch. The paths
that contribute to $\Delta_g$ must all go through the hub node $X$,
where the $m$ copies of $H_{g}$ are connected. Hence, to determine
$D_g$, all that is left is to calculate $\Delta_g$. The analytic
expression for $\Delta_g$, referred to as the crossing path
length~\cite{HiBe06,ZhGuDiChZh09}, can be derived as below.

Let $\Delta_g^{(\alpha,\beta)}$ be the sum of the lengths of all
shortest paths whose endpoints are in $H_{g}^{(\alpha)}$ and
$H_{g}^{(\beta)}$, respectively. According to whether the two
branches are one link long or two links long, we split the crossing
paths $\Delta_g^{(\alpha,\beta)}$ into two categories: the first
category composes of crossing paths $\Delta_g^{(1,\theta)}$ ($2 \leq
\theta \leq m$), while the second category consists of crossing
paths $\Delta_g^{(\theta_1,\theta_2)}$ with $2 \leq \theta_1 \leq
m$, $2 \leq \theta_2 \leq m$, and $\theta_1 \neq \theta_2$. It is
easy to see that the numbers of the two categories of crossing paths
are $m-1$ and $(m-1)(m-2)/2$, respectively. Moreover, any two
crossing paths in the same category have the same length. Thus, the
total sum $\Delta_g$ is given by
\begin{equation}\label{cross01}
\Delta_g =(m-1)\Delta_g^{(1,2)} +
\frac{(m-1)(m-2)}{2}\Delta_g^{(2,3)}\,.
\end{equation}
Having $\Delta_g$ in terms of the quantities of $\Delta_g^{(1,2)}$
and $\Delta_g^{(2,3)}$, the next step is to explicitly determine the
two quantities.

To calculate the crossing distance $\Delta_g^{(1,2)}$ and
$\Delta_g^{(2,3)}$, we give the following notation. For an arbitrary
node $v$ in network $H_{g}$, let $f_v(g)$ be the smallest value of
the shortest path length from $v$ to any of the $(m-1)^g$ peripheral
nodes belonging to $\mathbb{P}_g$, and the sum of $f_v(g)$ for all
nodes in $H_{g}$ is denoted by $F_g$. Analogously, in $H_{g}$ let
$h_v(g)$ denote the distance from a node $v$ to the hub node $X$,
and let $M_g$ stand for the total distance between all nodes in
$H_{g}$ and the hub node $X$ in $H_{g}$, including $X$ itself. By
definition, $F_{g+1}$ can be given by the sum
\begin{eqnarray}\label{bottom01}
F_{g+1} &=&\sum_{v\in H_{g}^{(1)}} f_v(g+1) +\sum_{\eta=
2}^{m}\sum_{v\in
H_{g}^{(\eta)}} f_v(g+1)\nonumber \\
&=& \sum_{v\in H_{g}}[h_v(g)+1]+(m-1)\,\sum_{v \in
H_{g}}f_v(g)\nonumber \\
&=&(m-1)\,F_{g}+N_g+M_g\,,
\end{eqnarray}
and $M_{g+1}$ can be written recursively as
\begin{eqnarray}\label{hub01}
M_{g+1} &=&\sum_{v\in H_{g}^{(1)}} h_v(g+1)+\sum_{\eta=
2}^{m}\sum_{v\in
H_{g}^{(\eta)}} h_v(g+1)\nonumber \\
&=& \sum_{v\in H_{g}}h_v(g)+(m-1)\,\sum_{v \in
H_{g}}[f_v(g)+1]\nonumber \\
&=&M_g+(m-1)(F_{g}+N_g)\,.
\end{eqnarray}
Using $N_g=m^g$, and considering $F_1=1$ and $M_1=m-1$, the
simultaneous equations~(\ref{bottom01}) and~(\ref{hub01}) can be
solved inductively to obtain:
\begin{equation}\label{bottom02}
F_{g} =m^{g-2}[(2g-1)m-2g+2]
\end{equation}
and
\begin{equation}\label{hub02}
M_{g} =m^{g-2}(m-1)(m+2g-2)\,.
\end{equation}

With above obtained results, we can determine $\Delta_g^{(1,2)}$ and
$\Delta_g^{(2,3)}$, which can be expressed in terms of these
explicitly determined quantities. By definition, $\Delta_g^{(1,2)}$
is given by the sum
\begin{eqnarray}\label{cross03}
\Delta_g^{(1,2)}&=& \sum_{u \in H_{g}^{(1)},\,v \in H_{g}^{(2)}} d_{uv}(g+1)\nonumber \\
 &=&
\sum_{u \in H_{g}^{(1)},\,v \in H_{g}^{(2)}}
\Big[h_u(g)+1+f_v(g)\Big]\nonumber \\
&=&\sum_{v \in H_{g}^{(2)}} \sum_{u \in H_{g}^{(1)}} h_u(g)+\sum_{u
\in H_{g}^{(1)}}\sum_{v \in H_{g}^{(2)}}[1+f_v(g) ]\nonumber
\\&=&N_g\,M_g+(N_g)^2+N_g\,F_g\,.
\end{eqnarray}
Inserting Eqs.~(\ref{bottom02}) and~(\ref{hub02})
into~(\ref{cross03}), we have
\begin{equation}\label{cross04}
\Delta_g^{(1,2)}=2\,m^{2g-4}\left[m^2+2(g-2)m-2g+4\right]\,.
\end{equation}
Proceeding similarly,
\begin{eqnarray}\label{cross05}
\Delta_g^{(2,3)}&=& \sum_{u \in H_{g}^{(2)},\,v \in H_{g}^{(3)}} d_{uv}(g+1)\nonumber \\
 &=&2\,[(N_g)^2+N_g\,F_g]\nonumber
\\&=&2\,m^{2g-4}\left[m^2+(2g-3)m-2g+4\right]\,.
\end{eqnarray}
Substituting Eqs.~(\ref{cross04}) and~(\ref{cross05})
into~(\ref{cross01}), we get
\begin{equation}\label{cross06}
\Delta_g=m^{2g-3}(m-1)^{2}(m+2g-2)\,.
\end{equation}
Substituting Eq.~(\ref{cross06}) into~(\ref{total01}) and using the
initial value $D_1=m(m-1)/2$, we can obtain the exact expression for
the total distance
\begin{equation}\label{total03}
D_g=\frac{1}{2}
m^{g-2}\left[4m+m^2-m^3+2m^{g+2}-4g\,m^g+(4g-6)\,m^{g+1}\right] \,.
\end{equation}
The expression provided by Eq.~(\ref{total03}) is consistent with
the result previously obtained~\cite{No03}. Then the analytic
expression for average distance can be obtained as
\begin{equation}\label{apl02}
 d_g =\frac{4m+m^2-m^3+2m^{g+2}-4g\,m^g+(4g-6)\,m^{g+1}}{m^2(m^g-1)} \,.
\end{equation}

%%%%%%%%%%%%%%%%%%%%%%%%%%%%%%%%%%%%%%%%%%%%%%%%
%% Figure 3
%%%%%%%%%%%%%%%%%%%%%%%%%%%%%%%%%%%%%%%%%%%%%%%%%
\begin{figure}
\begin{center}
\includegraphics[width=.35\linewidth,trim=100 10 100 0]{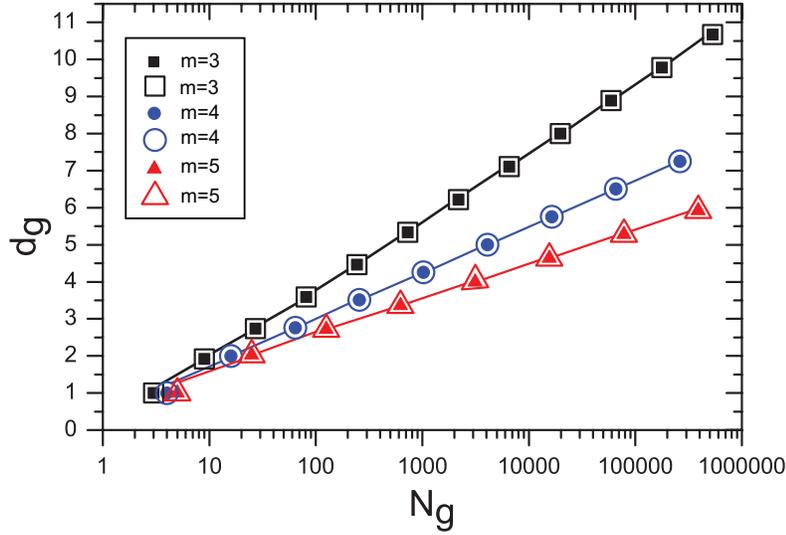}
\end{center}
\caption[kurzform]{\label{AveDis} Average distance $d_{g}$ versus
network order $N_g$ on a semi-logarithmic scale. The solid lines are
guides to the eye.}
\end{figure}
%%%%%%%%%%%%%%%%%%%%%%%%%%%%%%%%%%%%%%%%%%%%%%%%

We have also checked our rigorous result provided by
Eq.~(\ref{apl02}) against numerical calculations for different $m$
and various $g$. In all the cases we obtain a complete agreement
between our theoretical formula and the results of numerical
investigation, see figure~\ref{AveDis}.

We continue to express the average distance $d_g$ as a function of
network order $N_g$, in order to obtain the scaling between these
two quantities. Recalling that $N_g=m^g$, we have $g=\log_mN_g$.
Hence Eq.~(\ref{apl02}) can be rewritten as
\begin{eqnarray}\label{apl03}
 d_g &=&\frac{4m+m^2-m^3+4(m-1)N_g\log_mN_g+(2m^{2}-6m)N_g}{m^2\,(N_g-1)}\nonumber \\
 &=&\frac{4(m-1)N_g \ln N_g+(2m^{2}-6m) N_g \,\ln m+(4m+m^2-m^3)\ln m}{m^2\ln m\,(N_g-1)}\,.
\end{eqnarray}
In the infinite network order limit, i.e., $N_g\rightarrow \infty$
\begin{equation}\label{apl03}
 d_g =\frac{4(m-1)}{m^{2}\ln m}\ln N_g+2-\frac{6}{m}\,.
\end{equation}
Thus, for large networks, the leading behavior of average distance
grows logarithmically with increasing network order.

The above observed small-world phenomenon that the leading behavior
of average distance is a logarithmic function of network order can
be accounted for by the following heuristic arguments based on the
peculiar architecture of the networks. At first sight, this family
of modular networks is not a very compact system, since in these
networks, nodes with large degrees are not directly linked to one
another, but connected to those nodes with small degree. However,
this network family is made up of many small densely interconnected
clusters, which combine to form larger but less compact groups
connected by nodes with relatively high degrees. For node pairs in a
small group, their shortest path length is very small because of the
high cohesiveness of small modules. For the length of shortest paths
between two nodes belonging to different large groups, it seems long
because the groups that the nodes lie at are not adjacent to each
other. But this is not the fact. By construction, although the
relatively large groups are not directly adjacent, they are joined
by some large nodes, which are connected to each other by a layer of
intermediate small-degree nodes (see figure~\ref{network}), such as
the peripheral nodes or locally peripheral nodes~\cite{No03}. Thus,
different from conventional random scale-free networks, especially
assortative networks~\cite{Newman02}, in the studied scale-free
modular networks, although large-degree nodes are not connected to
one another, they play the role of bridges linking different modules
together, which is the main reason why the average distance of the
networks is small.

It deserves to be mentioned that, although the studied modular
scale-free networks display small-world behavior, the logarithmic
scaling of average distance with respect to network order is
different from the sublogarithmic scaling for conventional
non-modular stochastic scale-free networks with degree distribution
exponent $\gamma <3$, in which the average distance $d(N)$ behaves
as a double logarithmic scaling with network order $N$, namely,
$d(N)\sim \ln\ln N$~\cite{ChLu02,CoHa03}. Thus, despite that the
degree distribution exponent of the modular scale-free networks is
smaller than 3, their average distance is larger than that of their
random counterparts with the same network order. The root of this
difference may also lie with the modular structure, particularly the
indirect connection of large nodes, as addressed above. The genuine
reasons for this dissimilarity need further studies in the future.

\section{Conclusions}

The determination and analysis of average distance is important to
understand the complexity of and dynamic processes on complex
networks, which has been a subject of considerable interest within
the physics community. In this paper, we investigated analytically
the average distance in a class of deterministically growing
networks with scale-free behavior and modular structure, which exist
simultaneously in a plethora of real-life networks, such as social
and biological networks. Based on the self-similar structure of the
networks, we derived the closed-form expression for the average
distance. The obtained exact solution shows that for very large
networks, they are small-world with their average distance
increasing as a logarithmic function of network order. We confirmed
the rigorous solution by using extensive numerical simulations. We
also showed that the small-world behavior lies with the inherent
modularity and scale-free property of the networks.

\subsection*{Acknowledgment}

We would like to thank Xing Li for his support. This research was
supported by the National Natural Science Foundation of China under
Grants No. 60704044, No. 60873040, and No. 60873070, the National
Basic Research Program of China under Grant No. 2007CB310806,
Shanghai Leading Academic Discipline Project No. B114, the Program
for New Century Excellent Talents in University of China (Grants No.
NCET-06-0376), and Shanghai Committee of Science and Technology
(Grants No. 08DZ2271800 and No. 09DZ2272800).


\section*{References}
\begin{thebibliography}{ref1}


\bibitem{AlBa02}
R. Albert and A.-L. Barab\'asi, Rev. Mod. Phys. {\bf 74}, 47 (2002).

\bibitem{DoMe02}
S. N. Dorogovtsev and J. F. F. Mendes, Adv. Phys. {\bf 51}, 1079
(2002).

\bibitem{Ne03}
M. E. J. Newman, SIAM Rev. {\bf 45}, 167 (2003).

\bibitem{BoLaMoChHw06}
S. Boccaletti, V. Latora, Y. Moreno, M. Chavez and D.-U. Hwanga,
Phys. Rep. {\bf 424}, 175 (2006).

\bibitem{WaSt98} D. J. Watts and H. Strogatz,
       %Collective dynamics of `small-world' networks,
        Nature (London) {\bf 393}, 440 (1998).


\bibitem{ChLu02}
 F. Chung and L. Lu, Proc. Natl. Acad. Sci. U.S.A. {\bf 99}, 15879 (2002).

\bibitem{CoHa03}
R. Cohen and S. Havlin, Phys. Rev. Lett. {\bf 90}, 058701 (2003).

\bibitem{DoMeOl06}
S. N. Dorogovtsev, J. F. F. Mendes, and J. G. Oliveira, Phys. Rev. E
{\bf 73}, 056122 (2006).

\bibitem{SoHaMa06}
 C. Song, S. Havlin, H. A. Makse,
Nature Phys. {\bf 2}, 275 (2006).

\bibitem{ZhZhZo07}
Z. Z. Zhang, S. G. Zhou, and T. Zou, Eur. Phys. J. B {\bf 56}, 259
(2007).

\bibitem{ZhZhChGu08}
Z. Z. Zhang, S. G. Zhou, L. C. Chen, and J. H. Guan, Eur. Phys. J. B
{\bf 64}, 277 (2008).

\bibitem{XiMaiWaXiWa08}
Y. Xiao, B. D. MacArthur, H. Wang, M. Xiong, and W. Wang, Phys. Rev.
E {\bf 78}, 046102 (2008).

\bibitem{AdLuPuHu01}
L. A. Adamic, R. M. Lukose, A. R. Puniyani, and B. A. Huberman,
Phys. Rev. E {\bf 64}, 046135 (2001).

\bibitem{NiMoLaHo03}
T. Nishikawa, A. E. Motter, Y.-C. Lai, and F. C. Hoppensteadt, Phys.
Rev. Lett. {\bf 91}, 014101 (2003).

\bibitem{CoBeTeVoKl07}
S. Condamin, O. B\'enichou, V. Tejedor, R. Voituriez, and J.
Klafter, Nature (London) {\bf 450}, 77 (2007).

\bibitem{ZhLiZhWuGu09}
Z. Z. Zhang, Y. Lin, S. G. Zhou, B. Wu, and J. H. Guan, New J. Phys.
{\bf 11}, 103043 (2009).

\bibitem{ZhQiZhGaGu09}
Z. Z. Zhang, Y. Qi, S. G. Zhou, S. Y. Gao, and J. H. Guan, Phys.
Rev. E (in press). %{\bf 64}, 046135 (2001).

\bibitem{BaAl99}
A.-L. Barab\'asi and R. Albert,
      %Emergence of scaling in random networks,
       Science {\bf 286}, 509 (1999).

\bibitem{GiNe02}
M. Girvan and M. E. J. Newman, Proc. Natl. Acad. Sci. U.S.A. {\bf
99}, 7821 (2002).

\bibitem{RaSoMoOlBa02}
E. Ravasz, A. L. Somera, D. A. Mongru. Z. N. Oltvai, and A.-L.
Barab\'asi, Science {\bf 297}, 1551 (2002).

\bibitem{RaBa03}
E. Ravasz and A.-L. Barab\'asi, Phys. Rev. E {\bf 67}, 026112
(2003).

\bibitem{No03}
J. D. Noh, Phys. Rev. E {\bf 67}, 045103(R) (2003).

\bibitem{Zh03}
H. J. Zhou, Phys. Rev. E {\bf 67}, 061901 (2003).

\bibitem{NoRi04a}
J. D. Noh and H. Rieger, Phys. Rev. E {\bf 69}, 036111 (2004).

\bibitem{ZhLiGaZhGuLi09}
Z. Z. Zhang, Y. Lin, S. Y. Gao, S. G. Zhou, J. H. Guan, and M. Li,
Phys. Rev. E, in press.% {\bf 69}, 036111 (2004).J. Stat. Mech. (2009) P10022.

\bibitem{BaRaVi01}
A.-L. Barab\'asi, E. Ravasz, and T. Vicsek,
          %Deterministic scale-free networks,
          Physica A  {\bf 299}, 559 (2001).

\bibitem{IgYa05}
K. Iguchi and H. Yamada, Phys. Rev. E {\bf 71}, 036144 (2005).

\bibitem{AgBu09}
E. Agliari and R. Burioni, Phys. Rev. E {\bf 80}, 031125 (2009).

\bibitem{ZhLiGaZhGu09}
Z. Z. Zhang, Y. Lin, S. Y. Gao, S. G. Zhou,  and J. H. Guan, J.
Stat. Mech. (2009) P10022.

\bibitem{HiBe06}
M. Hinczewski and A. N. Berker, Phys. Rev. E {\bf 73}, 066126
(2006).

\bibitem{ZhGuDiChZh09}
Z. Z. Zhang, J. H. Guan, B. L. Ding, L. C. Chen, and S. G. Zhou, New
J. Phys. {\bf 11}, 083007 (2009).

\bibitem{Newman02}
M. E. J. Newman,
%Assortative mixing in networks,
Phys. Rev. Lett. {\bf 89}, 208701 (2002).


\end{thebibliography}
\end{document}